\documentclass[aps, prl, showpacs, twocolumn, amsfonts, amsmath, amssymb, superscriptaddress, floatfix]{revtex4-1}

\usepackage[T1]{fontenc}
\usepackage[latin9]{inputenc}
\usepackage{color}
\usepackage{graphicx}
\usepackage{dsfont}

\def\be{\begin{equation}}
\def\ee{\end{equation}}
\def\bea{\begin{eqnarray}}
\def\eea{\end{eqnarray}}

\def\esigma{E_{\sigma}}

\begin{document}

\title{Anderson transition of cold atoms with synthetic spin-orbit coupling in two-dimensional speckle potentials}

\author{Giuliano Orso} 
\affiliation{Universit\' e Paris Diderot, Sorbonne Paris Cit\' e,
Laboratoire Mat\' eriaux et Ph\'enom\`enes Quantiques, UMR 7162, 75013 Paris, France}



\begin{abstract}
We investigate the metal-insulator transition occurring in two dimensional (2D) systems of noninteracting atoms in the presence of artificial spin-orbit interactions  and a spatially correlated disorder generated by laser speckles.   
Based on a high order discretization scheme, we calculate the precise position of the mobility edge 
and verify that the transition belongs to the symplectic universality class.
We show  that the mobility edge depends strongly on the mixing angle between Rashba and Dresselhaus  spin-orbit couplings. For equal couplings a non-power-law divergence is found, signaling the crossing to the orthogonal class, where such a 2D transition is forbidden.
\end{abstract}

\maketitle

Anderson localization (AL)~\cite{Anderson:LocAnderson:PR58}, namely the absence of diffusion of a coherent wave in a disordered medium due to the
interference  between multiple-scattering paths, is a general phenomenon  observed for several kinds of waves, including 
 light waves in diffusive media~\cite{Wiersma:LightLoc:N97,Maret:AndersonTransLight:PRL06}
or photonic crystals~\cite{Segev:LocAnderson2DLight:N07,Lahini:AndersonLocNonlinPhotonicLattices:PRL08}, ultrasound~\cite{vanTiggelen:AndersonSound:NP08}, microwaves~\cite{Chabanov:StatisticalSignaturesPhotonLoc:N00} and atomic 
matter waves~\cite{Moore:AtomOpticsRealizationQKR:PRL95,Billy:AndersonBEC1D:N08,Roati:AubryAndreBEC1D:N08}, the latter describing the  behavior of atoms in the low-temperature quantum regime. 
  
Since AL finds its origin on interference effects, the space dimension 
as well as the symmetries of the model play a crucial role~\cite{Evers:AndersonTransitions:RMP08}.  
When both  spin-rotational  and time-reversal symmetries are preserved, the system belongs to
the \emph{orthogonal} universality class. 
While AL is the generic scenario in one and two dimensions,
in higher dimensions an Anderson phase transition occurs at a critical value of the energy $E=E_c$, called the mobility edge, 
separating localized states at lower energy from diffusive states at higher energy.
 
The inclusion of spin-orbit coupling (SOC) breaks SU(2) invariance and drives the system towards the  
\emph{symplectic} universality class. 
The spin of the particle rotates as the latter moves around a closed  loop and the direct and the time-reversed paths 
(on average) interfere destructively rather than constructively,  favoring diffusion rather than localization~\cite{HikamiSOCTwoDim:1980}. 
This spin-interference effect, called (weak) antilocalization,   has already been observed 
 for  2D electron gases in semiconductor quantum wells 
or at the surface of topological insulators. 
 A distinctive feature of the symplectic class is the occurrence of a 2D Anderson transition~\cite{Ando:SOCTwoDim:1989,FastenrathSOC:1991,Sheng:Rashba2Dlattice:2005} for strong SOC, but its experimental evidence is still lacking.
 




Ultracold atoms are natural candidates to fill the gap. Effects from atom-atom interaction can  be reduced via Feshbach resonances  and  a 
tunable random potential can be 
generated from laser speckles~\cite{SanchezPalencia:DisorderQGases:NP10}. Thus far experiments have focused on the orthogonal class. Recent achievements  
include the  observation~\cite{CoherentBackScatt:2012} of coherent backscattering in 2D systems and the 
study~\cite{Kondov:ThreeDimensionalAnderson:S11,Jendrzejewski:AndersonLoc3D:NP12,Semeghini:2014}  of the mobility edge
for the 3D Anderson transition. From the theoretical front, accurate numerical calculations~\cite{Delande:MobEdgeSpeckle:PRL2014, 
Pilati:LevelStats:2015,   Pasek:3DAndersonSpeckle:PRA2015,  Pilati:3DAndersonSpeckle:2015}  for $E_c$  have appeared going beyond  approximate 
estimates~\cite{Kuhn:Speckle:NJP07,Yedjour:MobilityEdge3D:EPJD10,Piraud:MobilityEdge3D:NJP13}. 
Atomic gases have also been employed to realize experimentally the quantum kicked rotor model and study AL
in momentum space. This setup has allowed a detailed investigation~\cite{Chabe:Anderson:PRL08,Lopez:ExperimentalTestOfUniversality:PRL12}  of the 3D 
Anderson transition and the observation of 2D AL ~\cite{Manai2DALkicked:2015}.
Parallel to these developments, significant  experimental and theoretical progress has been made
to create and control artificial SOC for cold atoms with the aim of exploring topological phases of quantum matter (for a review, see~\cite{Galitski:SpinOrbitNature:2013,Zhai:QuantumGasesWithSpinOrbit:2015}). 
Very recently, a synthetic SOC with \emph{tunable} Rashba~\cite{Bychkov:RashbaSpinOrbit:1984} and Dresselhaus~\cite{Dresselhaus:1955} 
terms has been experimentally 
realized~\cite{Huang2DSOC:2016} in 2D atomic gases,  
opening a new avenue to explore  Anderson transitions  in the symplectic class.

\begin{figure}
\includegraphics[width=0.98\columnwidth]{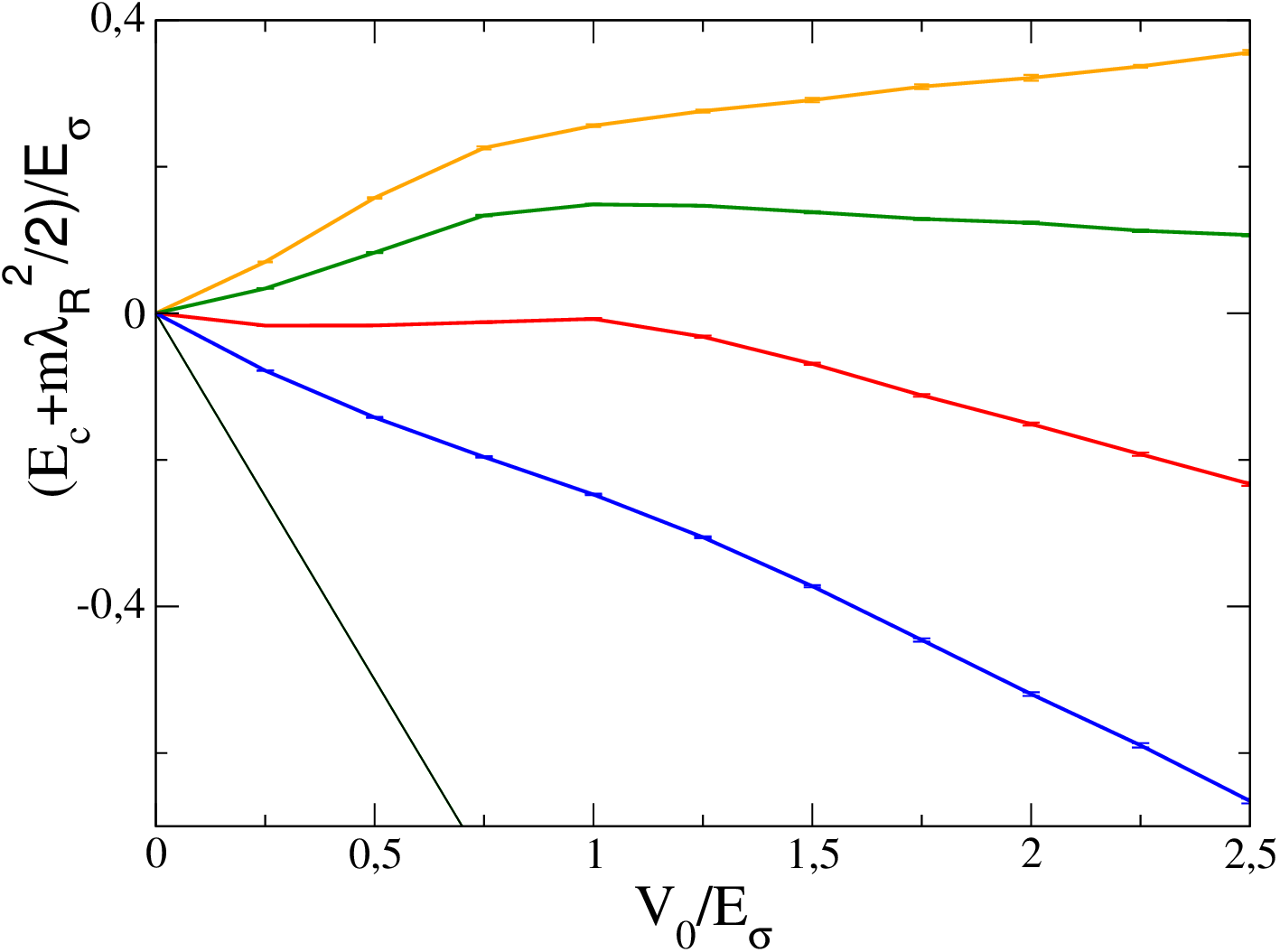}
\caption{(color online) Mobility edge $E_c$ of the 2D Anderson transition 
	separating low-energy localized states $(E<E_c)$  from high-energy
	diffusive states $(E>E_c)$, plotted
	as a function of the disorder amplitude $V_0$ and for increasing values of the 
	Rashba spin-orbit coupling  $\lambda_R m\sigma=0.03\; \textrm{(top curve)} , 0.1, 0.3$ and $ 1.256$, assuming $\lambda_D=0$.  
The black line corresponds to the energy bottom $E=-V_0-m\lambda_R^2/2$, below which no single particle state exists. Notice the linear 
behavior of $E_c$ for $V_0/E_\sigma\gtrsim 1$. }
\label{Fig:Mobility_Edge}
\end{figure}

\begin{figure}
	\includegraphics[width=0.95\columnwidth]{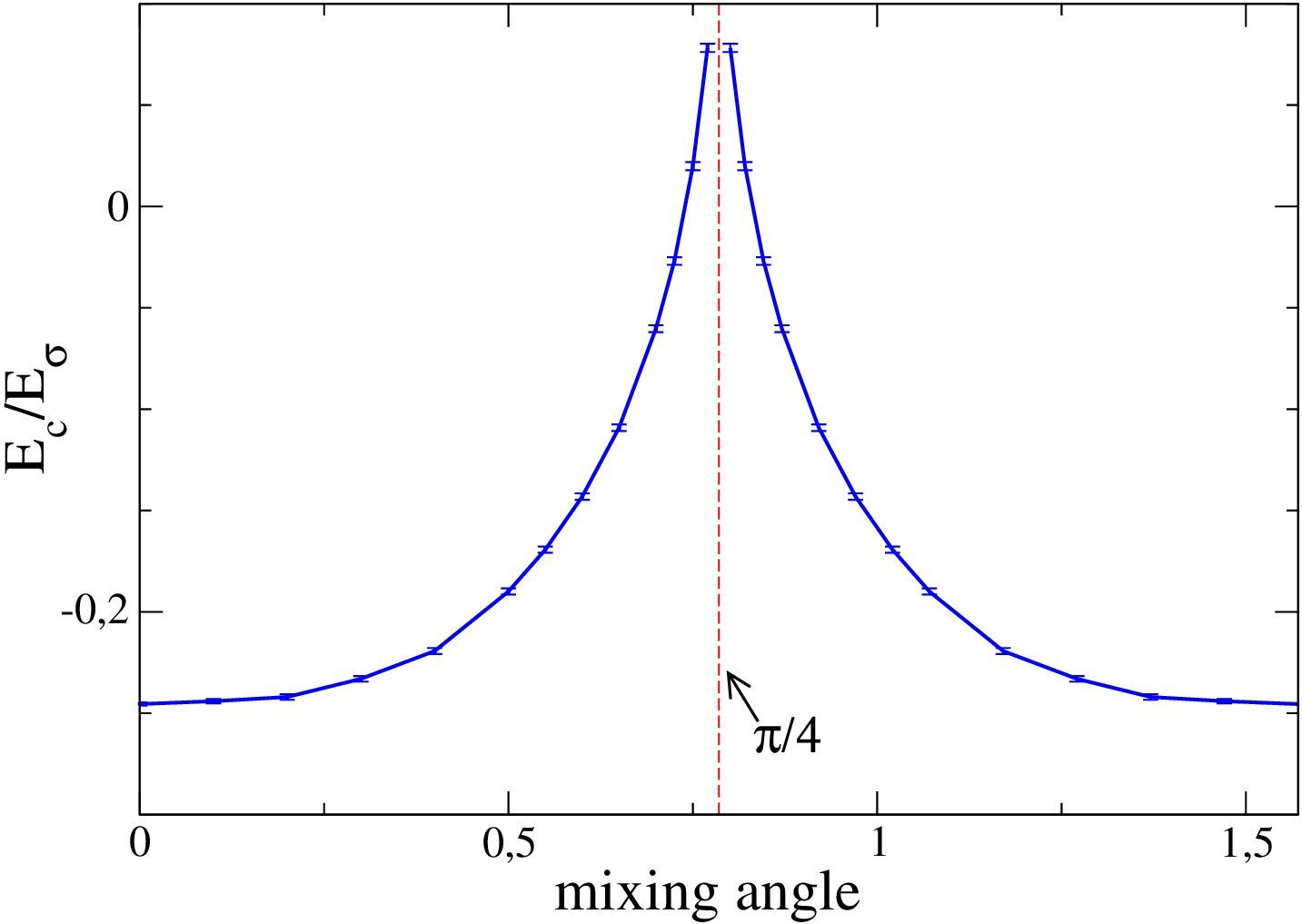}
	\caption{(color online) Mobility edge as a function of the mixing angle  $\theta=\arctan(\lambda_{D} /\lambda_{R})$ 
		between Rashba and Dresselhaus SOC, for a fixed value of the total strength $\sqrt{\lambda_R^2+\lambda_D^2}m\sigma=0.5$	
		and disorder amplitude $V_0=\esigma$. 
		Approaching $\theta=\pi/4$ (vertical dashed line), corresponding to equal strengths of Rashba and Dresselhaus SOC, the mobility edge rises sharply and actually diverges. Indeed, at this special point, spin scattering is absent and the model falls into
		 the orthogonal universality class, for which all states are localized in two dimensions implying $E_c=+\infty$. 
		 Notice that the mobility edge $E_c$ is invariant under the transformation 
		$\theta\rightarrow \pi/2-\theta$ 
		exchanging the Rashba and Dresselhaus terms in Eq.(\ref{Ham}).}
	\label{Fig:Dresselhaus}
\end{figure}

In this Letter we investigate the 2D Anderson transition in  atomic gases with artificial  SOC and subject to a laser speckle potential. We calculate numerically the precise position of the mobility edge, taking into full account the potential distribution and the spatial correlations of the disorder.
In particular: (i) we identify a regime where $E_c$ depends linearly on the disorder amplitude, with a slope 
decreasing and changing sign as the SOC increases (Fig.\ref{Fig:Mobility_Edge});  (ii) we show that the interference between Rashba and Dresselhauss SOC leads to a strong dependence of  $E_c$ on the mixing angle  with a non-power-law divergence
as the two magnitudes coincide (Fig.\ref{Fig:Dresselhaus}). Hence, by tuning the SOC one can induce an interesting crossover 
between symplectic and orthogonal universality classes. 

Previous theoretical studies of  atomic gases in the presence of both disorder and SOC have addressed AL  in
 1D quasiperiodic  lattices~\cite{AndersonLocSOCQuasiP:2012}, the dynamics of a 1D  Bose-Einstein condensate~\cite{Mardonov:GPwithSOC:2015}  and the competition between disorder and superfluidity in 2D Fermi gases~\cite{LiuBCSwithSOC:2016}.

The Hamiltonian of a spin-1/2 atom of mass $m$
in the presence of linear  SOC is given by:
\be\label{Ham}
H=\left( \frac{\mathbf k^2}{2m}+ V(\mathbf r)\right) \mathds{1}+\lambda_R (k_y \sigma_x-k_x \sigma_y)+\lambda_D(k_y \sigma_x+k_x \sigma_y),
\ee
where  $\mathbf k=-i\nabla$ is the momentum  of the particle (we use the convention $\hbar=1$) and $ V(\mathbf r)$ is the external speckle potential. Moreover
$\mathds{1}$ is the $2\times 2$ identity matrix,  $\sigma_x, \sigma_y$ are the Pauli matrices
and $\lambda_R$ and $\lambda_D$ correspond to the strengths of the Rashba and Dresselhaus couplings, respectively (for a discussion about realistic schemes to implement  such a model with cold atoms see Ref.~\cite{CampbellRDSOC:2011}). 
In the absence of disorder, a pure Rashba coupling yields split 
energy dispersions $E_{\mathbf k\pm}=k^2/2m\pm k\lambda_R$, with $k=|\mathbf k|$. The 
ground state occurs at $k=m \lambda_R$ with energy $-m \lambda_R^2/2$. 

In the following we shall focus on blue-detuned speckles, as employed in recent 
experiments~\cite{Kondov:ThreeDimensionalAnderson:S11,Jendrzejewski:AndersonLoc3D:NP12,Semeghini:2014}. Their  potential 
distribution follows the Rayleigh law~\cite{Kuhn:PRLSpeckle:2005,Goodman:07}:
\begin{equation}
 P(V) = \frac{\Theta(V+V_0)}{V_0} \exp{\left(- \frac{V+V_0}{V_0}\right)}
\label{Eq:Rayleigh-up}
\end{equation}
where  $\Theta$ is the Heaviside (unit step) function and $V_0$ is related to the variance by  $\langle V^2\rangle\!=\!V_0^2$. 
Notice that in Eq.(\ref{Eq:Rayleigh-up})  we have shifted the potential by its average value, without loss of generality. 

We generate the speckle potential  numerically by first computing the normalized electric field amplitude $ \epsilon(\mathbf r)$, whose real and imaginary parts are 
 normally distributed random variables with zero mean and unit variance. This quantity is then convoluted with the 
 point spread function  $h(\mathbf r)$ of the diffusive glass plate. Let us call $f(\mathbf r)$ the modulus square of the result;
  that is, $f(\mathbf r)=|\int d\mathbf r^\prime \epsilon( \mathbf r^\prime)h(\mathbf r-\mathbf r^\prime)|^2$. Then  the disorder potential is given by
  $V(\mathbf r)=V_0(f(\mathbf r)/f_\textrm{av}-1)$, where
$f_\textrm{av}=\int d\mathbf r f(\mathbf r)/S$ is the spatial average of $f$, with $S$ being the surface area.

 The spatial correlation function of the 2D speckle pattern can be written as 
$ \langle V(0)V(\mathbf{r}) \rangle=V_0^2
|h(\mathbf r)/h(0)|^2$. 
For a circular aperture,  the Fourier transform of
the point spread function is an Airy disk, $\tilde h(\mathbf k)=\Theta(k_0-|\mathbf k|)$, where 
$k_0=\alpha k_L$, $\alpha$  being the aperture angle and $k_L$ the wavevector of the laser beam. By using
 $\int_0^{2\pi} e^{-i k r \cos \theta} d\theta=2\pi J_0(kr)$ and $\int_0^{k_0} J_0(kr) k dk=J_1(k_0 r)k_0/r$, 
 $J_n(x)$ being the Bessel function of order $n$, we obtain 
\begin{equation}
 \langle V(0)V(\mathbf{r}) \rangle = V_0^2 \frac{4 J_1^2(r/\sigma)}{(r/\sigma)^2},
\label{Eq:Correlation_Speckle}
\end{equation}
where $\sigma=1/k_0$ is  the correlation length of the speckle pattern (see Supplemental Material 
\footnote{See Supplemental Material at 
http://link.aps.org/supplemental/10.1103/PhysRevLett.118.105301 for more
details on the numerical generation of the 2D speckle
potential and a thorough analysis of discretization effects
using the second and the fourth order discretization schemes.}). In the following we measure all energies in units of
the correlation energy $\esigma=1/(m\sigma^2)$.

\begin{figure}
\includegraphics[width=0.98\columnwidth]{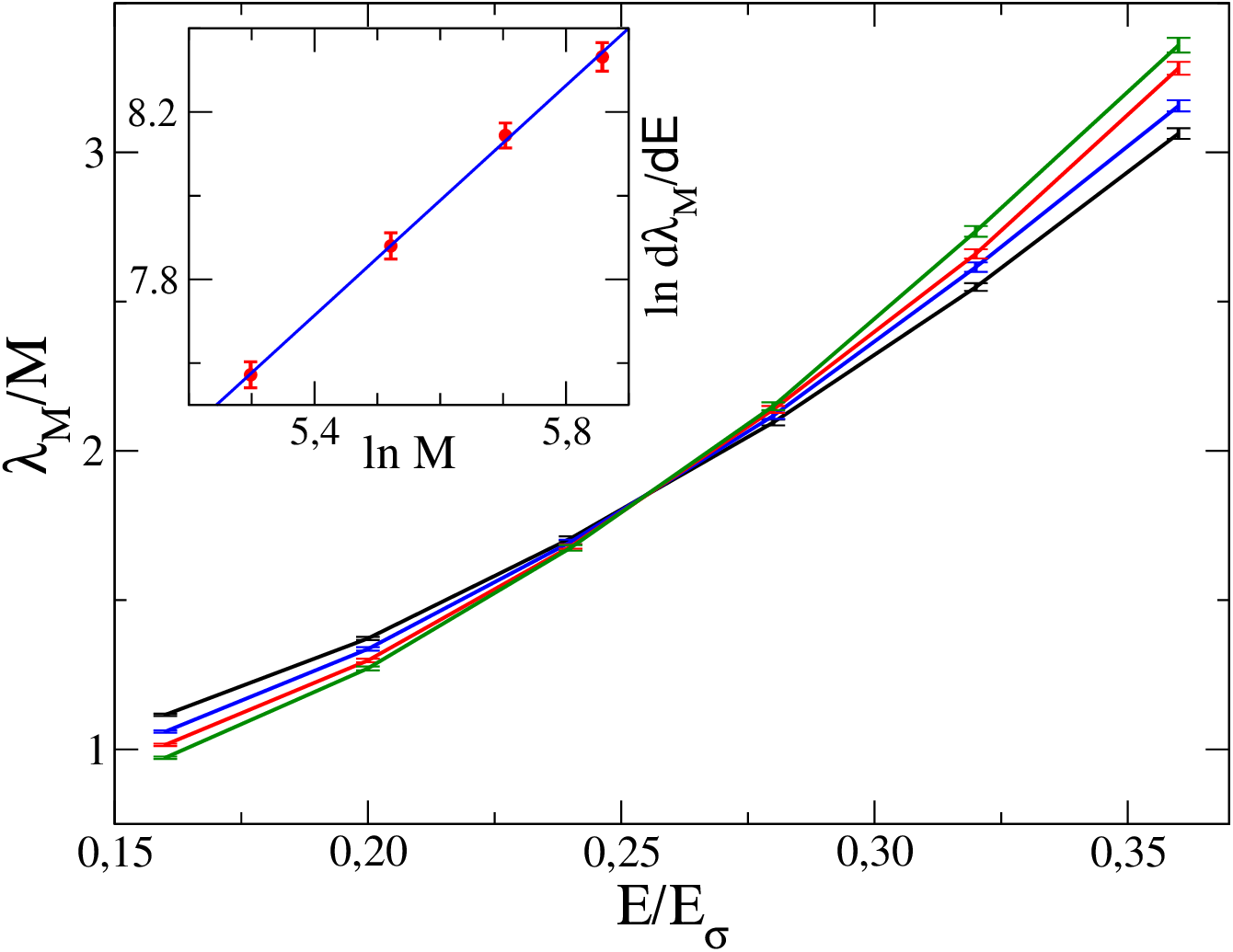}
\caption{(color online)  Numerical calculation of the critical point 
	 of the 2D Anderson transition for cold atoms with synthetic  Rashba SOC in a blue-detuned speckle. 
	 After discretization of the model on a strip-shaped grid of spacing $\Delta$, 
	 height $M$ and length $L\gg M$, we 
	 use the transfer-matrix method to calculate the localization length
	 $\lambda_M$.   
	 The main panel shows the ratio $\lambda_M/M$ as a function of energy 	 
	 calculated for increasing values of $M=200$ (top curve on the left), $250,300,350$ assuming $\Delta=0.2\pi \sigma$. The crossing point corresponds to the mobility edge, $E=E_c\simeq 0.256 \esigma$, with $\esigma=1/(m\sigma^2)$, $\sigma$ being the correlation length of the speckle; see Eq.(\ref{Eq:Correlation_Speckle}).
	 The Rashba  strength is $m\lambda_R \sigma=0.03 $ and the disorder amplitude is $V_0\!=\!\esigma$.	 
%
%
%
 The inset shows the evaluation of the critical exponent 
 $\nu$ from the scaling behavior of  $d\lambda_M/dE$ 
 at the mobility edge, see Eq.(\ref{fastnu}). }
\label{Fig:Crossing}
\end{figure}

In order to calculate the precise position of the mobility edge, we discretize the
stationary Schrodinger equation, $H\Psi=E\Psi$, on a grid by replacing first and second order derivates by finite differences.  Here $\Psi=(\psi_\uparrow \psi_\downarrow)^t$  is the two-component spinor wave function and $E$ is the energy of the particle.  
%
The simplest procedure, as employed   in  
  Ref.\cite{Delande:MobEdgeSpeckle:PRL2014}, is to use the second order central approximation: $\partial_x \psi_\sigma = (\psi_{i+1j\sigma}-\psi_{i-1j\sigma})/(2\Delta)+\mathcal{O}(\Delta^2)$
and $\partial_{xx} \psi_\sigma = (\psi_{i+1j\sigma}+\psi_{i-1j\sigma}-2\psi_{ij\sigma})/\Delta^2+\mathcal{O}(\Delta^2)$ (and analogously for the $y$ variable). 
Here  $\psi_{ij\sigma}\equiv \psi_\sigma(\mathbf r=\Delta i \mathbf e_x+ \Delta j \mathbf e_y)$, $\mathbf e_x$ and $\mathbf e_y$ being the unitary vector along the $x$ and $y$ axes, respectively, and $\Delta$  the discretization step. 
 This turned out to be unpractical for strong SOC, 
as it requires very fine grids to get converged  results 
for the transmission amplitude, 
increasing significantly the computational effort (see Supplemental Material). 
For this reason, we have used the fourth order approximation:
$\partial_x \psi_\sigma= (-\psi_{i+2j\sigma}+\psi_{i-2j\sigma} +8\psi_{i+1j\sigma}-8\psi_{i-1j\sigma} )/(12\Delta)+\mathcal{O}(\Delta^4)$
and $\partial_{xx} \psi_\sigma = (-\psi_{i+2j\sigma}-\psi_{i-2j\sigma} +16\psi_{i+1j\sigma}+16\psi_{i-1j\sigma} -30\psi_{ij\sigma})/(12\Delta^2)+\mathcal{O}(\Delta^4)$.

The retained scheme yields a generalized 2D Anderson model, where the
 coupling also extends to next-to-nearest neighboring sites.   
We consider a  strip-shaped grid with $L$ sites in the 
longitudinal direction and $M$ sites in the transverse one, with $M\!\ll\!L$. 
We also impose periodic boundary condition in the transverse direction to reduce finite-size effects. In this quasi-1D geometry,
 the system is  Anderson localized and we use the transfer matrix method~\cite{McKinnonKramer:TransferMatrix:ZPB83} 
 to accurately compute its  transmission amplitude $T$. For large $L$, the latter  decays exponentially  as 
 $T\propto \exp(-2L/\lambda_M)$, $\lambda_M$ being the 1D localization length.   

The critical point of the Anderson transition can be identified by 
calculating the ratio $\lambda_M/M$ as a function of energy and for increasing
values of $M$, as shown in Fig.~\ref{Fig:Crossing} (main panel). Here 
we have considered a pure Rashba  SOC with strength $\lambda_R m \sigma=0.03$ and disorder amplitude $V_0\!=\!\esigma$.
The grid spacing is  $\Delta=0.2\pi \sigma$ and $M$ varies between $200$ 
and $350$. Since the log of the total transmission is a self-averaging quantity,
we have calculated it  for grids of length $L=50000$ using $336$ different realizations of the disorder, 
and then averaging the obtained results. In this way the relative error in the 1D
localization length is below $0.7\%$.

 At low energy, in the localized regime, $\lambda_M$ converges to the 2D localization length $\xi=\lim_{M\to\infty} \lambda_M$
as $M$ becomes large, implying that the ratio $\lambda_M/M$ decreases with $M$.
In contrast, at high energy,  in the metallic phase, $\lambda_M/M$ increases with $M$,
whereas at the critical point, the ratio  takes  a (finite) constant value, $\lim_{M\rightarrow +\infty}\lambda_M/M=\Lambda_c$.
From the crossing point in Fig.~\ref{Fig:Crossing},  we find $E_c\simeq 0.256\esigma$ and $\Lambda_c\simeq 1.85$.

Next, we show that the 2D Anderson transition discussed here belongs to the symplectic  class. According to the one parameter scaling theory,
 the ratio $\lambda_M/M$ can be written in terms of a scaling function $f$ as
\begin{equation}\label{slevin1}
 \frac{\lambda_M(E)}{M} =f (u(\omega) M^{1/\nu}),
\end{equation}
where $u$ is a function of the reduced energy $\omega=(E-E_c)/E_c$ and
$\nu$ is the critical exponent. In Eq.(\ref{slevin1}) we have neglected possible contributions coming from 
irrelevant terms, since our values of $M$ are relatively large and no sizable 
drift of the crossing point is observed in Fig.~\ref{Fig:Crossing}. 

A first estimate of the critical exponent can be obtained by linearizing the
functions $f$ and $u$ in the proximity of the mobility edge. By substituting $f(x)=a_0+x$  and $u(\omega)=b_1 \omega$
in Eq.(\ref{slevin1}), where $a_0$ and $b_1$ are unknown constants, and taking the derivative of both sides with respect to the 
energy, we obtain that at the critical point 
\be\label{fastnu}
\frac{d\lambda_M}{dE}=\frac{b_1}{E_c}M^{1+1/\nu}.
\ee
We  calculate the  derivative in  Eq.(\ref{fastnu}) 
via central  difference using our numerical data at  $E=0.24\esigma$ and $E=0.28\esigma$, taking into account their statistical uncertainty. The  result is then plotted in the inset of Fig.~\ref{Fig:Crossing} as a function of $M$, using a log-log scale.    
 By fitting the data with a straight line of slope  $1+1/\nu$, we find $\nu=2.69\pm 0.21$, which is fully consistent with the best available~\cite{Asada:2Dsymplectic:PRL2002,Asada:BestEstimateSOC:2004} estimate $\nu=2.73 \pm 0.02$ for the 2D Anderson transition in the symplectic  class obtained in lattice models 
 with random SOC.

We can further improve the accuracy of our results by using the entire numerical data set.
For this purpose,  the functions $u$ and $f$ are Taylor expanded up to order 
$m$ and $n$, respectively, yielding $u(\omega)=\sum_{j=1}^{m} b_{j} \omega^{j}$
and $f(x)=\sum_{k=0}^n a_k x^k$, with $a_1=1$.  
The total number of fitting parameters is then given by  $2+m+n$.  
Following Ref.~\cite{Slevin:CriticalExponent:NJP14}, we perform a nonlinear least squares fit of the data, to extract the best estimates for the fitting parameters
and their error bars. With $n=m=3$ we obtain $E_c/\esigma=0.256\pm 0.002$, $\nu=2.67\pm 0.14$ and $a_0=\Lambda_c=1.855 \pm 0.02 $, corresponding to 
a reduced chi square $\chi_\textrm{red}=0.32$. 
Similar results can be found using smaller values of $m$ and $n$, by narrowing the fitting region  around the mobility edge.
Notice that, for fixed periodic boundary conditions and in the absence of discretization effects, $\Lambda_c$ is also  universal.  Our result compares well 
with the value  $\Lambda_c=1.844 \pm 0.002 $ obtained in Refs.~\cite{Asada:2Dsymplectic:PRL2002,Asada:BestEstimateSOC:2004}, suggesting that discretization effects are indeed rather small.

In Fig.~\ref{Fig:Mobility_Edge} we show the calculated mobility edge 
as a function of $V_0$  for increasing values of $\lambda_R$, going from $m\lambda_R \sigma=0.03$ (top curve) to $m\lambda_R \sigma=1.256$
(the inclusion of the Dresselhaus term will be discussed later). 
For vanishing SOC  and finite disorder strength all states are localized and   $E_c\rightarrow +\infty$.  
We see in Fig.~\ref{Fig:Mobility_Edge} that the mobility edge exhibits a kink
around $V_0\sim E_\sigma$ followed by an approximately linear behavior in the strong disorder regime, which is
reminiscent of classical percolation. Remarkably, the  slope depends  on the value of the Rashba SOC,
changing continuously from positive to negative values as $\lambda_R$ increases.
In contrast, for strong SOC,  $E_c$ always decreases  as $V_0$ increases, 
a situation already encountered for atoms in blue-detuned 3D laser speckles~\cite{Delande:MobEdgeSpeckle:PRL2014,Pilati:LevelStats:2015} without  SOC.

Notice that  discretization effects become more and more important as $\lambda_R$ increases 
(for $\lambda_R m\sigma=1.256$ we have used $\Delta=0.15\pi \sigma$).
Indeed the grid spacing must satisfy $\Delta\ll \textrm{min}(\sigma, \ell_\textrm{so})$, where 
$\ell_\textrm{so}=\pi/(m \lambda_R)$ is the spin-precession length.
For strong SOC, $\ell_\textrm{so}$ becomes the shortest length scale in the problem, implying that
very fine grids
are needed to accurately compute the position of the mobility edge.
Altogether, data shown in Fig.~\ref{Fig:Mobility_Edge} required
$700 000$ h of allocation time on a supercomputer
with $2$ Pflop/s.
  
 
Thus far, we have mainly focused on a pure  Rashba SOC by setting $\lambda_D=0$, but  the same results hold for a 
pure Dresselhaus SOC of the same strength. Indeed, the 
transformation 
$k_x\rightarrow -k_x$ in Eq.(\ref{Ham}) interchanges the 
Rashba and the Dresselhaus terms, leaving the total Hamiltonian invariant.
Let us now investigate the behavior of the mobility edge when   \emph{both} terms are present and interfere between each other
(weak antilocalization contributions to  conductivity from 
Rashba and Dresselhaus SOC are indeed not additive, see Ref.~\cite{Pikus:NoAntilocalization:1995}).  
 For this we write $\lambda_R=v_\textrm{so} \cos \theta$ and $\lambda_D=v_\textrm{so} \sin \theta$, where $v_\textrm{so}=\sqrt{\lambda_R^2+\lambda_D^2}$  and $\theta=\arctan(\lambda_D/\lambda_R)$ is the mixing angle. 
In Fig.\ref{Fig:Dresselhaus} we show the position of the mobility edge as a function of the mixing angle for  
$v_\textrm{so}=0.5/m\sigma$ and   $V_0=\esigma$.  Since $E_c$ is  invariant under the transformation 
$\theta\rightarrow \pi/2-\theta$, 
it is sufficient to study it for $\theta$ varying
between $0$ (pure Rashba) and $\pi/4$ (equal strengths of Rashba and Dresselhaus SOC). At $\theta=\pi/4$
the system is known~\cite{Bernevig:SpinHelixPRL:2006,Koralek:SpinHelixNature:2009} to exhibit an exact SU(2) symmetry, which generates persistent spin-helix states and is robust against spin-independent disorder.


We see in Fig.\ref{Fig:Dresselhaus} that the mobility edge is strongly dependent on the mixing angle and  diverges as $\theta$ 
approaches $\pi/4$. 
Indeed for  $\lambda_R=\lambda_D$
  the SOC term in Eq.(\ref{Ham}) reduces to $2 \lambda_R k_y \sigma_x$.
 Since $\sigma_x$ is Hermitian and commutes with $H$, we can find common eigenstates
for the two operators. Taking into account that the eigenvalues of $\sigma_x$
are $\epsilon_{\pm}=\pm 1$, the Hamiltonian decouples into two scalar sectors,
$H_\pm=(\mathbf k^2/2m+V(\mathbf r)) \pm 2 \lambda_R    k_y $,
implying that spin scattering is absent and
the 2D model belongs to the orthogonal class, for which all states are localized and $E_c=+\infty$.


A very interesting  and novel question concerns the nature (power law, logarithmic, etc.) of the divergence observed in Fig.\ref{Fig:Dresselhaus}. 
According to Wegner's theory~\cite{Wegner:CrossoverSymmetryBreaking:1986} (see also Ref.~\cite{Jung:CrossoverSymmetryBreaking:2016}),
which holds for quantum models in $2+\epsilon$ spatial dimensions, 
a small term breaking either the spin-rotational or the time-reversal symmetries induces a shift of the mobility edge which is a power law with exponent equal to $1/(2\nu_\textrm{orth})$, $\nu_\textrm{orth}$ being the critical exponent in the orthogonal class. 
In our 2D case $\nu_\textrm{orth}\rightarrow + \infty$,  so $E_c$ cannot diverge as a power-law of $|\lambda_R-\lambda_D|$.
The divergence is probably  \emph{logarithmic}, but proving it requires 
further  numerical  and/or analytical work.

In conclusion, we have shown that atoms with artificial Rashba and Dresselhaus SOC exposed to a 2D speckle potential
undergo an Anderson transition belonging to the symplectic universality class. We have computed the 
precise position of the mobility edge and identified a regime (Fig.~\ref{Fig:Mobility_Edge}) where the latter scales linearly as a function of  
the disorder strength, with a slope changing sign as the SOC increases.
Importantly, we have unveiled (Fig.~\ref{Fig:Dresselhaus}) that the mobility edge exhibits a non-power-law
divergence at the spin-helix point,  reflecting the crossover  to the orthogonal  class.
Our results call
for the extension of Wegner's theory~\cite{Wegner:CrossoverSymmetryBreaking:1986} to pure 2D systems, which by itself is a novel and interesting theoretical challenge.

Our  predictions can  already be  tested experimentally using ultracold atoms with tunable synthetic SOC.
Finally, we mention that the numerical approach developed here is completely general and can be applied to any kind of random potential, including  short range~\cite{Morong:2DPointlikedisorder:2015}.

We thank D. Delande and V. Josse for useful discussions. We also thank K. Slevin and T. Ohtsuki for correspondence and for drawing our attention to Refs.~\cite{Wegner:CrossoverSymmetryBreaking:1986, Jung:CrossoverSymmetryBreaking:2016}. This work was granted access to the HPC resources of TGCC under the allocations 2015-057301  and 2016-057629 made by GENCI (Grand Equipement National de Calcul Intensif).

\bibliographystyle{apsrev}
\bibliography{ArtDataBasev4}

\end{document}